# When Virtual Therapy and Art Meet:

A Case Study of Creative Drawing Game in Virtual Environments

LAUREN BARON
Computer and Information Sciences, University of Delaware

BRIAN COHN
Computer Science, University of Southern California

ROGHAYEH BARMAKI*
Computer and Information Sciences, University of Delaware
* Correspondence to: Roghayeh Barmaki, E-mail: rlb@udel.edu

There have been a resurge lately on virtual therapy and other virtual- and tele-medicine services due to the new normal of practicing "shelter at home". In this paper, we propose a creative drawing game for virtual therapy and investigate user's comfort and movement freedom in a pilot study. In a mixed-design study, healthy participants (N=16, 8 females) completed one of the easy or hard trajectories of the virtual therapy game in standing and seated arrangements using a virtual-reality headset. The results from participants' movement accuracy, task completion time and usability questionnaires indicate that participants had significant performance differences on two levels of the game based on its difficulty (between-subjects factor), but no difference in seated and standing configurations (within-subjects factor). Also hard mode was more favorable among participants. This work offers implications on virtual reality and 3D-interactive systems, with specific contributions to virtual therapy, and serious games for healthcare applications.

## 1 INTRODUCTION

Virtual Reality (VR) is a computer-generated simulation of a 3D environment that users can immerse themselves into and interact with via hardware (headset, controllers, joystick, treadmill, etc.). Given that most people in the world are experiencing stressful life changes under the COVID-19 pandemic crisis, we investigate how to integrate VR into at-home therapy. VR has been successfully used within rehabilitation settings for motor learning, impaired cognition, obesity, and overall health and wellness [9]. In this paper, we introduce a creative drawing game for virtual therapy and investigate user's comfort, range of motion and movement in multiple scenarios and configurations in a pilot study. This game allows the user to be fully engaged in both the physical stimulation and the mental stimulation. Figure 1 demonstrates an overview of the game with a user performing one of the therapeutic tasks while standing. The game encourages broad arm motions while still being entertaining as the user strives to connect the dots of the drawing. A creative drawing game modelled like connect the dots allows for familiarity and ease while playing – users are not overwhelmed with a game that feels foreign to them. The working hypothesis of this study was that our creative drawing VR game would be effective when integrated into therapy by analyzing improved Task Completion Time (TCT), accuracy based on lower number of the mistakes, and user experience (UX). More specifically, research questions to inspire the study were: Does the VR therapy game improve user's range of motion and reach? Is there any difference between the complexities of the easy/hard levels in the game based on TCT and accuracy? How about and differences on TCT and accuracy based on game configurations of seated and standing? Do our users enjoy playing the game and recommend it to their peers? If so, which configurations are the most popular ones?

Figure 1: Overview of a user playing the creative VR therapy game.

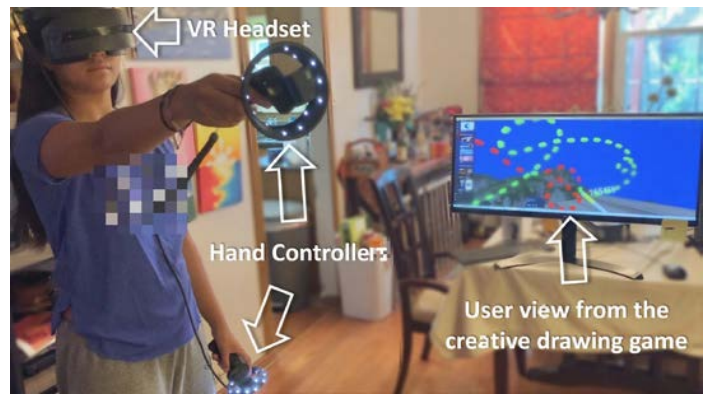

## 1.1 Motivation

Though VR started out as a form of entertainment, it has grown to have implications in the medical field, from simulating surgery for surgeons [22] to attenuating patient pain during chemotherapy [17]. One of the industries VR is advancing is physical therapy. VR is preferred in rehabilitation because of its portability so that patients can take the therapy home, its ability to attenuate pain, its independence from external pressures and distractions, and its game-like characteristics that engage users. For example, VR was proven to be better at reducing phantom limb pain than other distraction methods [7]. Because the patients used VR exercise imagery, it stimulated the same brain regions that are responsible for actual movement. Therefore, pain was reduced due to pain distraction and punctually activated brain regions involved in the pain matrix network [10]. Another example is how stroke patients report physical pain and an inability to concentrate during their rehabilitation without VR [24, 25]. Not only does VR make therapy movements not as painful, but it also helps patients regain movement that was lost (i.e. after a stroke) and extend their range of movement. By transforming rehabilitation into an entertaining game, the intense, repetitive, task-oriented arm exercises become more engaging and provide a more positive experience for both the patient and their therapist [11]. VR improves movement range and pain in the upper extremities (UE). Our paper looks at the difference between seated and standing VR regarding the dynamics of movement and comfort in a creative VR therapy game.

The pros and cons of seated VR and standing VR have been studied closely, yet still stimulate further research and discussion [27]. There are many reasons why people chose seated VR over standing VR and vice versa. Some users prefer to be seated at a desk because they can have an interactive surface (desk) to perform their task on [1]; it is more comfortable and less prone to fatigue to be seated than walking around during long durations [1]; it is more suitable for those with a sedentary lifestyle or mobility-impairments [26]; it reduces the risk of injuries due to falls, motion sickness, or hitting other objects [26]; and it makes them users feel less vulnerable and more acceptable to use VR [27]. However, standing VR gives users better range for full-body gestures, better performance, and better interactions and locomotion within the 3D environment [26]. In addition, developers lean towards seated VR because hand/object tracking can be easier when the user's overall movements are restricted in space [1]; many leading VR products appear to be designed for seated configuration [1]; it is more suitable for small or cluttered spaces [26]; and it is less likely for users to be entangled in the chords/cables [28]. There are many numerous hardware and layout configurations for VR, and there is still a lot of progress to be made towards best practices of user comfort, and movement assessment of users in seated VR, particularly in virtual therapy. We aim to evaluate what configuration is best for virtual therapy for UE mobility by measuring accuracy and TCT, visualizing their movements through data tracking, and evaluating their responses to the exit survey.

Investigating the best practices for VR therapy is in high demand. The coronavirus (COVID-19) pandemic is a global health emergency currently involving 188 countries with >29,145,000 infections confirmed and >925,000 deaths worldwide [13]. However, COVID-19 has affected nearly every person mentally and emotionally. The impact on mental health concerns not only medical staff, who are working nonstop in a high-stress and high-risk health environment, but also millions of people forced into isolation/quarantine [9, 12]. Availability of physical therapy services in the community—even for urgent concerns—has decreased during the COVID-19 pandemic, as opinions about whether home- and community-based physical therapy (PT) should remain open are mixed [8]. Because of stay-at-home mandates, patients must choose to take their PT home or risk exposure going to one of the scarce PT services that are open amidst the pandemic. By continuing their PT at home, patients reduce the risk of hospitalization or other forms of care—both essential public health goals during a viral pandemic that is currently overwhelming hospital and nursing home capacity [8]. It is necessary to increase remote access to care while preserving scarce resources, including personal protective equipment [16]. Without integrating remote rehabilitation options, such as VR, telehealth services, and digital practices, practitioners may disproportionately harm the most vulnerable patients, send a troubling message to the general public about the value of physical therapists, or worsen the potential short-term and long-term mental health consequences related to this global emergency [8, 12].

**1.2 Related Work**

We address the problem of how configuration contributes to user performance and range of motion in the VR environment. Our studies were inspired by several previous experiments that investigated the significance of body position while using VR. There are several VR systems based on different user body configurations: seated [4], leaning while seated [15], standing [4], leaning while standing [15], walking in place [21], etc. Kruijff et al. studied how leaning configurations affect VR performance [15]. They tested both static leaning (keeping a tilted posture throughout the whole trial) and dynamic leaning (their upper-body inclination changes dynamically throughout the trial) with the leaning angles of forward, upright, and backwards. While the dynamic leaning data did not produce substantial results, the static leaning showed that leaning does improve accuracy, TCT, and range of movement. Their conclusions were based off the positive effect that leaning while seated had on self-motion perception. Self-motion perception is how users use sensory cues to be immersed in their VR environment; it increases task performance because users feel part of the virtual environment and perceive cues that anchor them to the real world [15, 28].

Range of motion while using VR is important to study because if we can find the best configuration for users to move freely in, their user performance and comfortability while completing the task will improve. Also, being able to use your body to navigate within the virtual environment allows your hands to be free to complete tasks. For instance, using 2D devices (joystick, keyboard, etc.) to move around is not practical. In most applications of VR, ground navigation is not the primary action the user has to perform, so the system should keep users' hands available to use for tasks other than ground navigation [20]. By moving in the virtual environment using other body parts, the user's hands, eyes, and local head orientation are completely free and available for other physical or social interactions. LazyNav looked at how moving both the upper and lower extremities affected user performance [20]. 30 participants tested combinations of these motions while standing: bend bust, lean bust, rotate shoulders, rotate hips, bend hips, bend knees, take a step. One-way ANOVA tests showed that there were significant differences of all the motion pairs when they measured their movement distance, TCT, and accuracy. With their quantitative and qualitative data, they were able to suggest which general body motions were easier to perform and more comfortable for their "lazy" VR design. This shows how a user's range of motion affects how they perform in their task and how easy/comfortable they perceive the game. When we evaluate how seated versus standing configurations, we will measure both user performance and user range of motion with quantitative (movement distance, TCT, accuracy, exit survey data) and qualitative data (user suggestions).

From gathering data on what configuration allows for the best range of motion during a VR game, we can propose the best configuration a patient should be in for VR physical therapy. One of the biggest advantages of using VR for PT is that it is portable; patients can take their therapy home to do it frequently at their convenience [11]. A virtual reality therapy home-based system (VRT-Home) was developed for children with hemiplegic cerebral palsy (CP) to practice hemiplegic hand and arm movements; children with CP have a brain injury in the motor cortex that impairs the opposite UE [18]. Their results showed that the system successfully targeted hand/arm movements of the hemiplegic UE, especially reaching activities that involve the shoulder and elbow. Additionally, the child participants reported "[having] lots of fun" and "would like to take the games practice therapy activities home to play". Patients enjoy using VR in their PT for their upper limbs; it is effective, enjoyable, and portable. However, this study only used seated configurations for their participants. We look at seated and standing configurations for a therapy game that works primarily on UE rehabilitation.

We also consider how the therapy game will be received as a remote therapy tool during the COVID-19 pandemic and "shelter at home". The Secret Garden is a 10-minute self-help VR protocol made to reduce the burden of the coronavirus [12]. VR is an effective tool for the prevention and treatment of stress-related psychopathological symptoms and PTSD, with therapeutic benefits [3, 19]. In this simulation, each user has a partner to discuss their emotions/reflections and they perform different tasks related to personal identity and interpersonal relationships. It provides the sense of community that was taken away due to isolation and quarantine and provides an outlet to manage one's stress. Its biggest takeaways were the flexible use, high level of autonomy, and lower costs. Gao et al. explores how physical activity VR programs reduce stress and promote health and wellbeing in older adults [9]. Our creative drawing game combines UE movements with a creative release to enhance user experience and health.

## 2  MATERIALS AND METHODS

The goal of this study is to compare two configurations of seated versus standing VR in body movements, user-friendliness, comfort and immersion during a creative drawing VR therapy game. Our creative drawing game was coded in C# and through Unity game engine. Our game is compatible with several VR Head-Mounted Displays (HMDs), but we have chosen Windows MR for our study as a more portable candidate in comparison to other consumer-level HMDs. Another motivation for choosing Windows MR is its easy setup which will be extremely important in the future for tele-rehabilitation use of the game at homes by ordinary, first-time VR users.

In our animal drawing game, the user can choose whether they want to participate in the Easy Level or Hard Level. The goal is to connect the dots of an outline drawing of either a fish (Easy Level) or a chicken (Hard Level) with a virtual paint brush. The background is a simple, serene mountain scape with a blue sky and clouds, allowing the user to focus on their task in a relaxing, distraction-free environment. When each dot is hit, it turns green from red, and a positive audio feedback sound is played to the user. When all the dots are green, meaning the user successfully connected all the dots of the drawing, they are celebrated by visually exciting animations. A walkthrough of each of the levels, based on their completion state is provided in Figure 2.

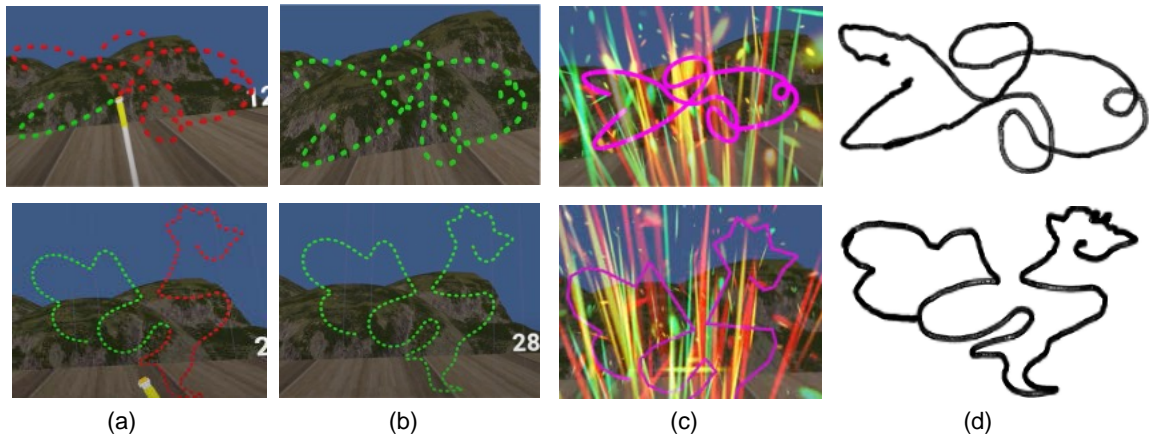

(a)　　　　　　　　(b)　　　　　　　　(c)　　　　　　　　(d)

Figure 2: A walkthrough of the game for each level (the Easy Level is the fish drawing and the Hard Level is the chicken drawing): (a) the dots are initially red until the user traces the drawing with the paintbrush, making the dots green; (b) the drawing is complete when all the dots are green; (c) the drawing comes to life in an animation and the user is celebrated with fireworks when the task is complete; (d) data visualization from the virtual paintbrush (hand controller) while drawing as they were drawing.

The game performance is flexible; the user can switch controllers to draw with either their left or right hand. The controller that is not drawing can be used to adjust the 3D dots model to the height or position the user feels most comfortable and allows this game to be played both seated and standing. No matter where they adjust the task to be, they are still reaching and moving their body to complete their drawing. We want to compare how much the range users can reach and how accurate their movements are while seated and while standing.

## 2.1 Questionnaires

Using Qualtrics, we developed pre- and post-questionnaires. In the pre-questionnaire, all the participants were asked about their demographics, prior VR and video game experience, level of education, past injuries, and fitness level. After the participants completed their tasks, they were given their individualized link to the post-questionnaire based on what testing group they were in. We chose to use a five-point Likert Scale for our questions because it is one of the most fundamental and frequently used psychometric tools for research [14]. They were all asked to rate their discomfort level, movement restrictions, and ease of completing the task using the Likert Scale. We then asked a variety of questions about their UX. Some of our post-questionnaire questions were derived from a validated and unified UX questionnaire for Immersive Virtual Environments [23].

        The pre-questionnaire allowed us to collect demographics and background data on our participant pool. We asked about demographics: age, gender, height, weight, ethnicity, and education level. This data helps us understand how diverse and representative the group we are collecting data from is. Participants were also questioned for VR usability: How often do you play video games; How much do you enjoy playing video games; Have you ever used Virtual Reality headsets before. For those who have experience with VR, we then asked: What is the reason you used VR; Did you enjoy using VR. This information indicates how receptive users will be to use a VR game for therapeutic activities. We also asked questions about user fitness: How frequently do you exercise a minimum of 20 minutes per session; Have you had a severe upper body injury, either due to sports or other accidents. Data about users' fitness tells us how well users' performances will be in our therapy game that targets the upper body. Users who have had previous rehabilitation experience would be able to provide more insight on how our therapy game compares in its effectiveness and entertainment.

        The post-questionnaire provides insight on VR usability/UX. Users were asked how strongly they agree to the following statements: It was easy to compete the virtual drawing task; I felt comfortable while completing the task; I felt my movement was restricted while completing the task; Using the VR drawing activity, I did stretch my body more than I normally do; I enjoyed playing the creative drawing game. This data gives us a better sense of how well our game will be perceived as a creative therapy game for users with little VR experience. We also asked about presence and cybersickness to ensure the users were not distracted by external stimuli while completing the task in the VR environment: The sense of moving around inside the virtual environment was compelling; My interactions with the virtual environment seemed natural; I was completely captivated by the virtual environment; I still paid attention to the real environment; I suffered from fatigue, headaches, nausea, or eyestrain during my interaction with the virtual environment. Learning that the user was immersed in the VR environment and was not distracted by external factors helps us eliminate that confounding variables contributed to their TCT and number of mistakes while playing. We can also assess how engaged users were with their therapeutic activities. Finally, we asked users how likely they were to recommend this creative therapy game to friends of family members as a therapeutic exercise, particularly amidst COVID-19. We also collected text entries from participants, asking for thoughts on how to improve this activity for future use and their preferences for being seated vs. standing. A unique ID was generated by each participant and was repeatedly used in completion of the questionnaires to assist us keep track of their pre-, post, and main intervention data while preserving their anonymity.

## 2.2 Study Design and Procedure

**Design** We chose to use a 2x2 counter-balanced mixed-design to test seated VR vs. standing VR and Easy Level vs. Hard Level. We used a within-subject design for the seated vs standing configuration because it allows us to see how the same person responds differently given the different conditions of seated vs. standing, and used a between-subject design for the level complexities because it reduces confounding variables due to exposure of multiple treatments. The four testing groups following the mixed design that participants were randomly assigned to were as follows: users test the easy level seated first then standing; users test the easy level standing first then seated; users test the hard level seated first then standing; users test the hard level standing first then seated.

**Procedure** The study was approved by the Institutional Review Board (Protocol #: XXXXXX). After consent, participants were randomly assigned to one of the four study conditions. We carefully instructed each participant what to do and after they expressed that they fully understood, they were guided to a chair or the middle of the room to stand depending on their conditions. We then gave them the headset to put on and their hand controllers and were started on either the Easy or Hard level using their dominant hands (Figure 3). We manually recorded how many times they made a mistake in their continuous drawing stroke (when a dot doesn't turn green because they missed it) and their task completion time. They were then asked to complete the post-questionnaire for the configuration they just tested and repeated the task in the other configuration. After all their trials were completed, each participant filled out the exit survey about their experience using the game.

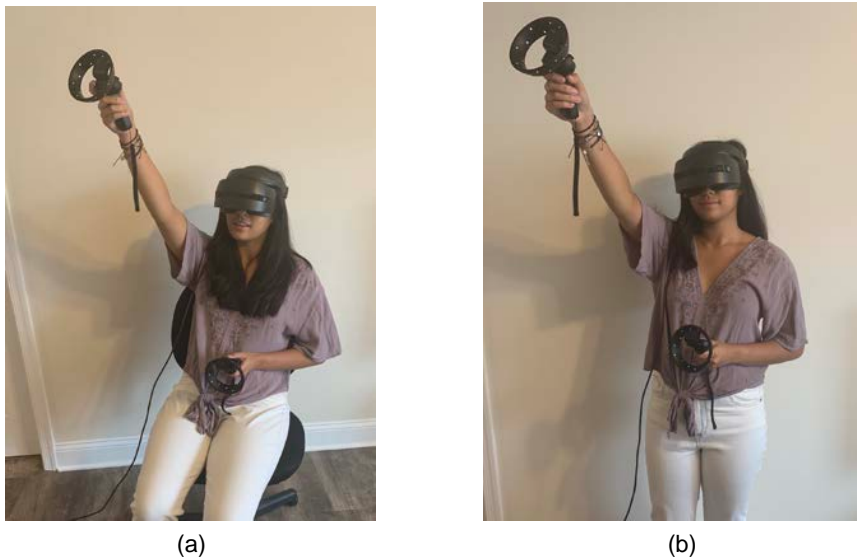

(a)          (b)

Figure 3: Study setup of the different configurations tested: (a) user is seated; (b) user is standing. User has on the required hardware: VR HMD and VR Hand controllers

**Participants** We assembled a participant pool of 16 (8 female) healthy volunteers enrolled in an undergraduate program for preliminary testing. Participants' age ranged from 18 to 20 (M = 19.06, SD ±0.56). 62.5% (10/16) of participants have Caucasian or White ethnicity, 31.25% (5/10) of participants came from an Asian or Pacific Islander background, and 6.25% (1/16) was Hispanic or Latino. Only 18.75% (3/16) have previously experienced a severe upper body injury, either due to sports or other incidents; 6.25% (1/16) needed to participate in rehabilitation sessions to recover from an upper body injury.

## 2.3 Statistical Analysis and Evaluation of Study Design

We used STATA software and RStudio to perform the setst and visualize the findings. We performed One-way ANOVA tests to find a significant difference between the Easy and Hard Levels based on our dependent variables of task completion time (TCT) and number of mistakes. We also performed tests to find out any significant differences between the seated and standing configurations using the TCT, number of mistakes, and postquestionnaire data. We also used a Python program to collect data on the hand controller position, which we visualized it using RStudio. The 3D visualization allows us to visually assess how accurate was a drawing compared to the provided outline. It also visualizes the range of their movements from how far they could reach to draw their drawings so that we can see what areas they struggled to reach. For example, in Figure 2d, the visualization shows that the user was shakiest in the upper left corner of the fish and overlapping lines in the upper left corner of the chicken; we can speculate that the user's range of motion is not as strong on the upper left side. The quantitative and qualitative data we collected from the questionnaires help us determine if their performance was affected by any confounding variables, such as distractions or cybersickness, and helps us understand their impressions about the game.

## 3 RESULTS

The results from 16 participants (8 females) took part in the pilot study is reported here. A significant difference between the game's level of complexity for Easy Level (fish drawing) and the Hard Level (chicken drawing) with task completion time ($F_{(1,30)}=8.52$, $p<0.01$), and number of mistakes was observed ($F_{(1,30)}=48.89$, $p<0.0001$). Overall, the chicken model was more challenging, and harder to complete, but yet entertaining; users also spent more time to complete it, and have more mistakes while performing the task. No significant difference was observed on study configurations of seated and standing based on TCT nor mistakes in any of the Easy or Hard Levels of the game. Figures 4 and 5 presents a summary of these results in four smaller conditions, and also for larger levels of the game based on TCT and mistakes.

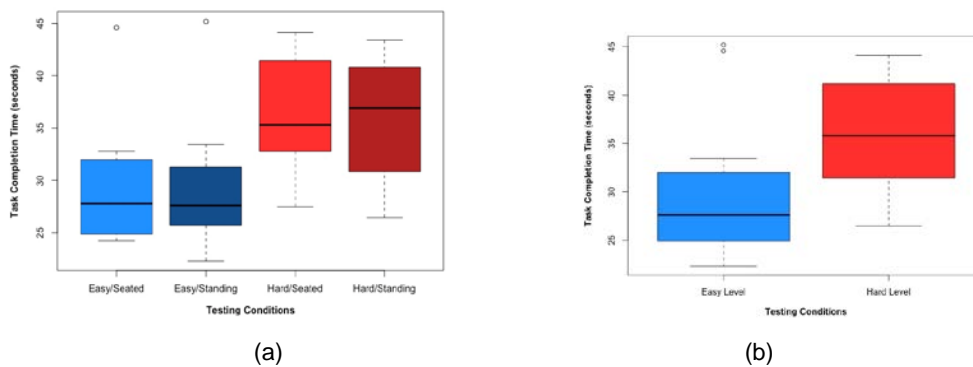

(a) (b)

Figure 4: Boxplots for task completion time (in seconds): (a) four study conditions (b) two larger conditions of easy and hard modes regardless of study configuration. Both of the plots indicate significant difference. Lower values are better.

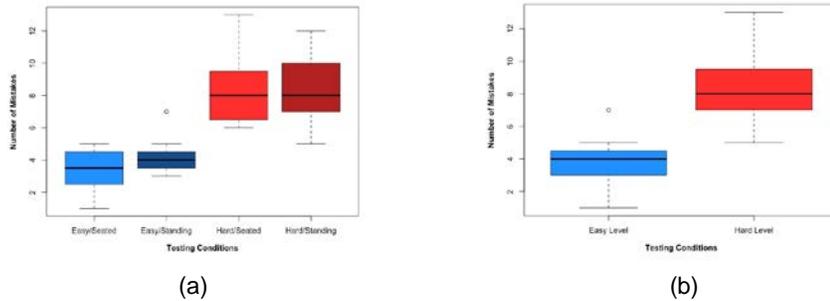

(a)　　　　　　　　　　　　　　(b)

Figure 5: Boxplots for number of mistakes: (a) four study conditions (b) two larger conditions of easy and hard modes regardless of study configuration. Both of the plots indicate significant difference. Lower values are better.

Moreover, we found that despite it's difficulty, participants *enjoyed* completing the chicken level more than fish level ($F_{(1,30)}$=10.91, *p<0.005*), and also they were more likely to *recommend* the game to a friend/family for therapy when played the chicken level ($F_{(1,30)}$=4.50, *p<0.05*). Participants also reported they were more *comfortable* to complete the task while seated ($F_{(1,30)}$=4.51, *p<0.05*), plus they also reported in the questionnaires that they have *stretched more* while seated ($F_{(1,30)}$=6.26, *p<0.01*). Other measures listed in the questionnaires were found to be less significant with study conditions and variables.

Moreover, many participants provided their overall feedback in the exit survey. For example:

> *"When I broke my arm from hockey and got my cast off, I had to do several exercises like bending/straightening my arms, rotating my wrists, and other things to regain my strength. I could see how this game could be helpful and fun at the same time".*

> "I prefer standing. It was a lot easier to move. When I was sitting I had to reach more and wanted to lift myself off the chair a bit to get to the highest parts of the chicken. But since I couldn't I had to really stretch my arm and controller out to hit the dots."

In summary, seated configuration was reported to be more comfortable, it was recognized to facilitate more stretching/reaching because they could not move lower body towards dots, and it had more accommodates to medical conditions. Besides, standing configuration was preferred by some of the participants because it was easier to reach while standing, especially for short participants. Another reason for seated is that some participants felt they had to be more weary of hitting things in the real world while standing because they were not grounded on a chair.

## 4　DISCUSSION

To evaluate the functionality of our portable creative therapy game, we tested our game in the apartment of a lab member, following COVID-19 guidelines. This setting strengthens our evidence that our therapy game is compatible with remote therapy and the shelter-at-home mandate. With our well-defined study design, we managed to conduct a relatively comprehensive data collection. However, similar to any study, there was some limitations to our pilot study. This was our preliminary study to objectively evaluate our VR therapy game and thus, we tested it with convenience sampling and our pool of participants was young, healthy college students. We anticipate testing the creative therapy game with actual patients in need of upper extremity therapy in the future.

This project contributes not only to upper limb rehabilitation, but to therapy of all disciplines. If we can effectively integrate VR into PT, patients will be more comfortable and more engaged in their recovery. Intensive, repetitive, task-oriented effective recovery tasks that are proven to be the most effective form of PT can be in the form of an entertaining, immersive VR game, from creative drawing for the upper body to versions of soccer to work on footwork [2, 5, 11]. Our project provides evidence that we can make a simple, familiar, portable game that is also helpful, encouraging, and distraction-free for remote therapy. Due to "Stay at Home" orders and need to socially distance in response to COVID-19, portable physical therapy is necessary. The demand for remote rehabilitation is high, as meeting with a physical therapist would increase the chance of infection and hospitalization [8]. Not to mention, those not suffering from UE mobility inhibitions are now feeling an increase of depression, stress, and anxiety from the global emergency [12]. Our proposed therapy game allows users to do their therapy at the comfort of their homes, helping both physical and emotional health of the individuals.

The immediate next steps of this project are to conduct studies with larger-scaled, and representative participant pools to validate our findings. Because this game would be targeted to UE patients, we would need participants with upper limb impairments to collect data from. From a technical point of view, the therapy game and data-collection software need to be more user-friendly for patients with no technical background to easily use at home. It would benefit the patient if they could receive real time feedback on their performance and compare it to previous performances on one easy-to-use interface. Also, using a creative physical therapy game requires the user to play the levels multiple times a day for the therapy to be truly effective [11]. Therefore, we need to add more levels so that the game stays challenging, engaging and entertaining for the user. More levels would also allow us to address the needs of multiple patients and market to not just stroke, Parkinson's disease or CP patients, but to all patients with upper extremity impairments.

## 5  CONCLUSION

In this article, a creative virtual therapy game was introduced and tested with a preliminary participant pool of students. The results from participants' movement accuracy, task completion time and usability questionnaires indicate that participants had significant performance differences on two levels of the game based on its difficulty (between-subjects factor), but no difference was observed in seated and standing configurations (within-subjects factor). It means that both of these configurations can be used interchangeably, for instance in future clinical applications with some further considerations; without introducing the risk of lowered performance or accuracy due to study configuration. Also hard mode was more popular among participants. These findings suggest great potential for its future applications in remote physical therapy for upper-extremity mobility during the COVID-19 shelter-at-home reality.

**ACKNOWLEDGMENTS**

We would like to acknowledge and thank all members in of the research team at the affiliated University. We also wish to express our gratitude to the participants who kindly participated in our study. We wish to acknowledge our gratitude for the 'Sponsor' program for sponsoring this project, with a grant from the 'GrantSponsor'. Any opinions, findings, and conclusions or recommendations expressed in this material are those of the authors and do not necessarily reflect the views of the sponsors.

[22] Tako, Y., Geri, A.Y., Avisar, M. and Teichman, E. 2017. Surgical Navigation Inside A Body. US20170367771A1. Dec. 28, 2017.
[23] Tcha-Tokey, K., CHRISTMANN, O., Loup-Escande, E. and Richir, S. 2016. Proposition and Validation of a Questionnaire to Measure the User Experience in Immersive Virtual Environments. *International Journal of Virtual Reality*. 16, 1 (2016), 33–48.
[24] Tistad, M., Tham, K., von Koch, L. and Ytterberg, C. 2012. Unfulfilled rehabilitation needs and dissatisfaction with care 12 months after a stroke: an explorative observational study. *BMC Neurology*. 12, 1 (Jun. 2012).
[25] Turville, M.L., Walker, J., Blennerhassett, J.M. and Carey, L.M. 2019. Experiences of Upper Limb Somatosensory Retraining in Persons With Stroke: An Interpretative Phenomenological Analysis. *Frontiers in Neuroscience*. 13, (Jul. 2019). DOI:https://doi.org/10.3389/fnins.2019.00756.
[26] Xu, W., Liang, H.-N., He, Q., Li, X., Yu, K. and Chen, Y. 2020. Results and Guidelines From a Repeated-Measures Design Experiment Comparing Standing and Seated Full-Body Gesture-Based Immersive Virtual Reality Exergames: Within-Subjects Evaluation. *JMIR Serious Games*. 8, 3 (2020), e17972. DOI:https://doi.org/10.2196/17972.
[27] Zielasko, D. and Bernhard, E.R. Sitting vs. Standing in VR: Towards a Systematic Classification of Challenges and (Dis)Advantages. *ResearchGate*.
[28] Zielasko, D., Weyers, B., Bellgardt, M., Pick, S., Meibner, A., Vierjahn, T. and Kuhlen, T.W. 2017. Remain seated: towards fully-immersive desktop VR. *2017 IEEE 3rd Workshop on Everyday Virtual Reality (WEVR)* (Mar. 2017), 1– 6.